# Water Oxidation and Hydrogen Evolution with Organic Photooxidants: A Theoretical Perspective

W. Domcke[1] and A. L. Sobolewski[2]

[1] Department of Chemistry, Technical University of Munich, D-85747 Garching, Germany
E-mail: domcke@ch.tum.de

[2] Institute of Physics, Polish Academy of Sciences, PL-02-668 Warsaw, Poland
 Email: sobola@ifpan.edu.pl

**ABSTRACT**

In this perspective, we discuss a novel water-splitting scenario, namely the direct oxidation of water molecules by organic photooxidants in hydrogen-bonded chromophore-water complexes. In comparison with the established scenario of semiconductor-based water splitting, the distance of electron transfer processes is thereby reduced from mesoscopic scales to the Ångstrøm scale and the time scale is reduced from milliseconds to femtoseconds, which suppresses competing loss processes. The concept is illustrated by computational studies for the heptazine-$H_2O$ complex. The excited-state landscape of this complex has been characterized with *ab initio* electronic-structure methods and the proton-coupled electron-transfer dynamics has been explored with nonadiabatic dynamics simulations. A unique feature of the heptazine chromophore is the existence of a low-lying and exceptionally long-lived $^1\pi\pi^*$ state in which a substantial part of the photon energy can be stored for hundreds of nanoseconds and is available for the oxidation of water molecules. The calculations reveal that the absorption spectra and the photochemical functionalities of heptazine chromophores can be systematically tailored by chemical substitution. The options of harvesting hydrogen and the problems posed by the high reactivity of OH radicals are discussed.

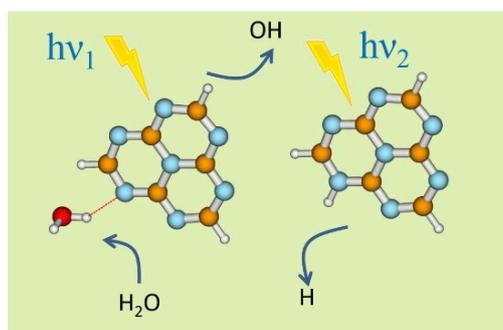

# 1. INTRODUCTION

The generation of hydrogen by electrolysis of water with renewable electricity is by now a mature technology, although the need of large amounts of precious metals is a serious bottleneck for global deployment. Solar water splitting, i.e. the direct decomposition of water into molecular hydrogen and molecular oxygen with photocatalysts, could be a more efficient way to convert solar energy into renewable chemical energy. Fifty years ago, the seminal paper of Fujishima and Honda stimulated the vision of solar-driven decomposition of water into molecular hydrogen in an electrochemical device with $TiO_2$ and Pt electrodes.[1]

The experiment of Fujishima and Honda and its interpretation have intrigued researchers through the past five decades. Photoelectrochemical water splitting with $TiO_2$ or related transition-metal (TM) oxides has extensively been investigated and the basic mechanisms seem to be well understood.[2-5] However, the band gap of TM oxide semiconductors generally is too large for efficient harvesting of solar energy. By a major effort in materials science, semiconductors with significantly lower band gaps have been identified and explored as photocatalytic materials for water splitting.[6-8] Unfortunately, the chemical stability of most of these semiconductors under the operating conditions of photoelectrochemical devices is currently too low for technological applications.[9,10] The lack of photostability also is a serious unsolved problem with the organometallic TM-metal complexes which are used as co-catalysts for hydrogen evolution or oxygen evolution.[11,12]

Unexpectedly and counterintuitively, a metal-free material, polymeric carbon nitride (PCN), has been found to be photocatalytically active for hydrogen evolution with near-UV light and to exhibit remarkable long-term photostability.[13] This material of nominal composition $C_6N_9H_3$ was first prepared by Berzelius in 1834 and termed "melon" by Justus von Liebig[14]. PCN materials, also known as graphitic carbon nitrides (g-$C_3N_4$),[15] are readily obtained from nitrogen-rich small molecules by pyrolysis.[16] The decent efficiency of PCNs for hydrogen evolution from sacrificial electron donors and their high photostability triggered a vast research effort during the past decade. There exists an immense literature on photocatalytic hydrogen evolution with polymeric carbon nitrides. We refer to refs.[17-19] for representative reviews. The PCNs are considered as semiconducting materials by the community and the goal has been the improvement of the efficiency of exciton dissociation into charge carriers as well as the improvement of charge carrier mobility and lifetime. However, the poorly constrained chemical compositions and molecular structures of these polymeric materials have been detrimental to explorations of the precise energy conversion mechanisms.[20]

In this perspective, we summarize recent insights which were obtained by taking a deliberately complementary approach toward water splitting which focusses on the spectroscopic and photochemical properties of the molecular building blocks of PCNs rather than on their semiconducting properties as solid materials. In particular, our analysis focusses on the heptazine (heptaazaphenalene) molecule which is the building block of g-$C_3N_4$.[15] A derivative of heptazine (Hz), trianisole-heptazine (TAHz), has recently been synthesized as a purified molecule and its photocatalytic activity was explored in homogeneous solution.[21] TAHz in organic (toluene) solution was found to be photochemically exceptionally stable and to exhibit remarkable spectroscopic properties. TAHz in aqueous suspension, on the other hand, was found to be able to photooxidize water (which was confirmed by the detection of OH radicals) and to evolve molecular hydrogen with an efficiency which is comparable to polymeric g-$C_3N_4$.[21] On the theoretical side, the excited electronic states of Hz and derivatives thereof were characterized with accurate ab initio calculations. The photochemical reaction mechanisms in hydrogen-bonded complexes of Hz or TAHz with substrate molecules were explored by calculations of excited-state reaction paths and potential-energy (PE) surfaces as well as by simulations of the ultrafast photochemical dynamics with nonadiabatic trajectory calculations.[22-24] Based on this extensive set of experimental and computational data, we suggest a molecular scenario of water oxidation and hydrogen evolution which is fundamentally different from the established scenario based on semiconductor physics.

## 2. WATER-SPLITTING PHOTOCATALYSIS WITH SEMICONDUCTORS

The standard free energy difference for the splitting of water into $H_2$ and $O_2$ is 113.5 kcal/mol, which corresponds to 1.23 eV per photon (four photons are needed to drive the four-electron $O_2$ evolution process). In semiconductor photocatalysis, a photon with an energy in excess of the band gap generates charge carriers (electrons in the conduction band and holes in the valence band). The charge carriers are presumed to separate and migrate to the solid-liquid interface,[25] where they react with adsorbed water molecules.[1] It has been shown that water molecules adsorbed on nano-scale $TiO_2$ particles are partially heterolytically dissociated into protons and $OH^-$ anions.[26, 27] Electrons arriving at the interface thus can neutralize the protons and holes can neutralize the $OH^-$ anions.

For $TiO_2$ and related semiconductors, this scenario is supported by extensive experimental data for charge carrier concentrations, charge carrier mobilities as well as rate constants for redox reactions at interfaces.[2, 4, 5, 28-30] Charge carrier lifetimes have been determined to be in the range of picoseconds to nanoseconds, while the timescales of interfacial charge transfer and redox reactions were estimated to be in the range of milliseconds to seconds.[28, 29] There is thus a mismatch of several orders of magnitude between the lifetime of the charge carriers and the timescales of productive redox reactions. It has been estimated that enhancements of the charge carrier lifetimes by factors of $10^3 - 10^9$ would be required for efficient energy conversion in semiconductor photocatalysis.[2] The expectations that co-catalysts could speed up the surface redox reactions by orders of magnitude did not materialize in recent decades despite extensive research.[5, 7, 31] The main bottleneck is the oxygen evolution reaction, requiring the accumulation of four holes, which is kinetically extremely challenging. Most experiments actually consider only one of the two half-reactions (hydrogen evolution or oxygen evolution) and are performed with sacrificial substrates which serve as convenient electron or hole scavengers.

For these reasons, the realization of a technologically feasible, scalable, and economically viable process based on light-driven water splitting with semiconductor materials has remained elusive.[32] The yields of molecular hydrogen and molecular oxygen achieved with the photoelectrochemical water-splitting scheme have remained disappointingly low. Although the improvement of the efficiency of water splitting with semiconductor photocatalysts is generally perceived as a problem of materials science, the available data on charge carrier generation, transport, and reactivity in $TiO_2$ indicate that the core of the problem is the photoelectrochemical scenario itself, namely the "indirect" oxidation of water. In particular, the thermodynamically and kinetically most challenging reaction step, the oxidation of water, is spatially (by micrometers to millimeters) as well as temporally (by milliseconds to seconds) separated from the energy harvesting event (the absorption of the photon). While the initial steps (photon absorption and charge separation) are very fast, the final reaction steps (redox reactions with water) are slow. Large losses due to charge carrier recombination and trapping therefore appear inevitable. Moreover, the surviving charge carriers have to home in on rather dilute reaction partners, such as $H_3O^+$ and $OH^-$, at the solid-liquid interface, which implies a significant kinetic challenge.

## 3. ALTERNATIVE SCENARIO: "DIRECT" WATER OXIDATION WITH ORGANIC PHOTOOXIDANTS

Although the basic mechanistic challenges of photoelectrochemical water splitting are well documented and well understood, there have been few attempts to explore alternative scenarios. The concept of artificial photosynthesis with metal-organic donor-chromophore-acceptor complexes[33-35] also invokes charge separation followed by water oxidation, albeit for shorter distances than in semiconductor photocatalysis. In this perspective, we discuss an alternative scenario: the direct light-driven oxidation of water molecules with the strongest available organic photooxidants in pre-associated hydrogen-bonded complexes of the substrate with the chromophore. We suggest that the direct oxidation of water via a photoinduced reaction in pre-assembled chromophore-water complexes may represent a mechanistically superior alternative to the indirect photoelectrochemical water-splitting scenario described above.

We postulate a chromophore A which is a base (proton acceptor) in the electronic ground state and a strong oxidant (electron acceptor) in the photoexcited state. Being a base, A spontaneously associates with water molecules by hydrogen bonding, forming A⋯HOH complexes (the dots denote hydrogen bonding). The site of the deposition of the photon energy is thus as close as possible to the water molecule to be oxidized. The distance of electron transfer is reduced from mesoscopic scales to the Ångstrøm scale. To be a good absorber in the visible or near-UV spectrum, A likely is an aromatic molecule, that is, an aromatic base. The initial step of the photoinduced reaction is

$$A \cdots HOH \ + \ h\nu \ \rightarrow \ A^* \cdots HOH. \tag{1}$$

where A* denotes the chromophore in a bright excited state, typically a $^1\pi\pi^*$ state. If A* is a sufficiently strong oxidant, it may be able to abstract an electron from the water molecule

$$A^* \cdots HOH \rightarrow A^- \cdots HOH^+. \qquad (2)$$

The electronic charge separation generates a strong driving force for the transfer of a proton from the water molecule to the chromophore (the proton wants to follow the electron), resulting in charge neutralization

$$A^- \cdots HOH^+ \rightarrow AH \cdots OH. \qquad (3)$$

This process has been termed "electron-driven proton transfer" (EDPT)[36] and is more widely known as excited-state proton-coupled electron transfer (PCET).[37] Notably, the AH and OH radicals in Eq. (3) are formed in their electronic ground states. Part of the energy of the absorbed photon is thus stored as chemical energy in the radical pair.

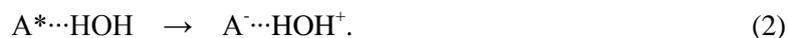

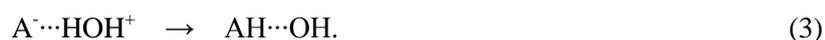

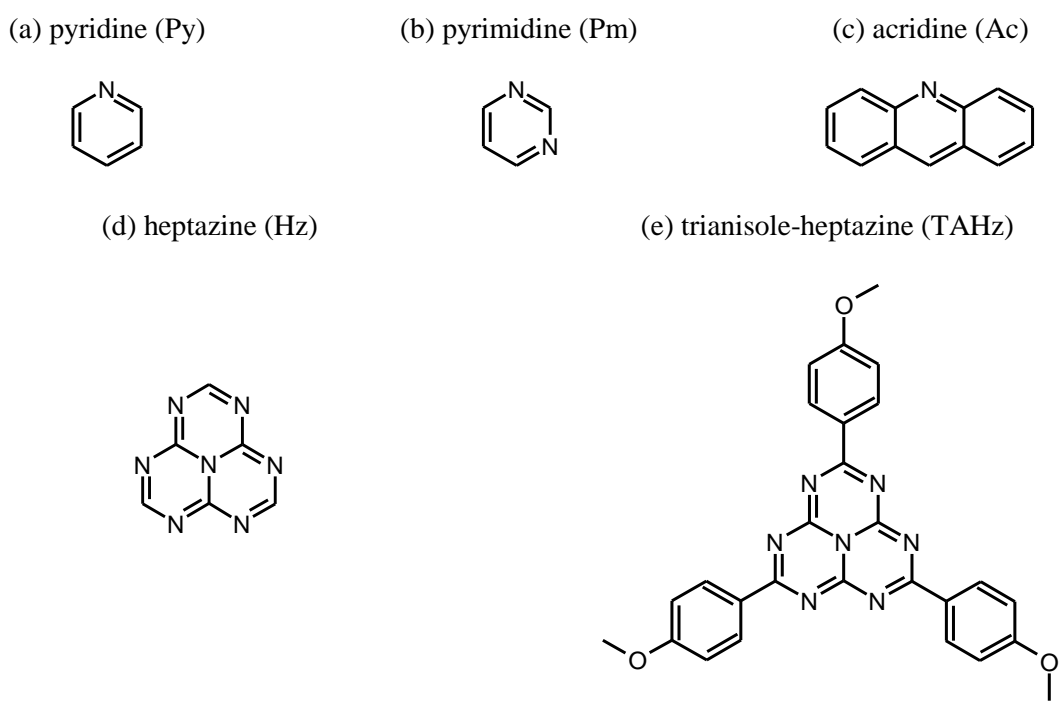

*Fig. 1. Chemical structures of pyridine (a), pyrimidine (b), acridine (c), heptazine (d), and trianisole-heptazine (e).*

The simplest aromatic base is pyridine (Py), see Fig. 1a. Its lowest singlet and triplet excitation energies are listed in Table 1. It has been shown in a molecular beam experiment under very precisely defined conditions that Py excited to its lowest $^1\pi\pi^*$ excited state can oxidize water in Py-$(H_2O)_n$ clusters with n > 3.[38] The photoinduced reaction yields PyH and OH$(H_2O)_{n-1}$ radicals. The PyH radical was detected via its characteristic ionization potential and its electronic absorption spectrum. The reaction mechanism was confirmed by ab initio calculations of excited-state PE surfaces and by nonadiabatic molecular dynamics simulations.[39-41] Analogous experiments and calculations were performed for clusters of pyrimidine (Pm, see Fig. 1b) with water.[42] These studies demonstrate that water molecules can be homolytically dissociated with photon energies which are significantly below the photodissociation threshold of water (6.66 eV or 186 nm). Because the absorption bands of Py and Pm (see Table 1) are far outside the solar spectrum at the surface of Earth, Py and Pm are, like $TiO_2$, model systems for the exploration of the fundamental mechanisms of the photooxidation of water rather than practically useful photocatalysts.

Table 1. Vertical excitation energies and oscillator strengths (in parentheses) of the three lowest singlet excited states and the three lowest triplet excited states of pyridine, pyrimidine, acridine, heptazine and trianisole-heptazine, calculated with the ADC(2) method and the cc-pVDZ basis set at MP2-optimized ground-state equilibrium geometries.

| Py | | Pm | | Ac | | Hz | | TAHz | |
|---|---|---|---|---|---|---|---|---|---|
| state | ΔE/eV | state | ΔE/eV | state | ΔE/eV | state | ΔE/eV | state | ΔE/eV |
| $^1\pi\pi^*$ | 5.14(0.004) | $^1\pi\pi^*$ | 4.49(0.005) | $^1\pi\pi^*$ | 3.76(0.09) | $^1\pi\pi^*$ | 2.57(0.0) | $^1\pi\pi^*$ | 2.62(0.0) |
| $^1\pi\pi^*$ | 5.35(0.025) | $^1n\pi^*$ | 4.80(0.0) | $^1\pi\pi^*$ | 3.90(0.05) | $^1n\pi^*$ | 3.76 | $^1\pi\pi^*$ | 3.62(2.38) |
| $^1n\pi^*$ | 5.37(0.0) | $^1\pi\pi^*$ | 5.53(0.23) | $^1n\pi^*$ | 3.92(0.0) | $^1\pi\pi^*$ | 4.43(0.54) | $^1n\pi^*$ | 3.66(0.0) |
| $^3\pi\pi^*$ | 4.43 | $^3n\pi^*$ | 4.10 | $^3\pi\pi^*$ | 2.43 | $^3\pi\pi^*$ | 2.85 | $^3\pi\pi^*$ | 2.87 |
| $^3n\pi^*$ | 4.54 | $^3n\pi^*$ | 4.61 | $^3n\pi^*$ | 3.65 | $^3\pi\pi^*$ | 3.67 | $^3\pi\pi^*$ | 3.28 |
| $^3\pi\pi^*$ | 5.04 | $^3\pi\pi^*$ | 4.66 | $^3\pi\pi^*$ | 3.68 | $^3n\pi^*$ | 3.76 | $^3n\pi^*$ | 3.66 |

Obviously, the absorption bands can be lowered in energy by extension of the size of the aromatic frame. An example is 5-methoxyquinoline, which according to a recent simulation is able to oxidize water by a photoinduced PCET reaction.[43] Another example is acridine (Ac, see Fig. 1c). Ac exhibits a $^1\pi\pi^*$ excitation energy in the near UV (3.5 eV), see Table 1. Calculations predict that the barrier for PCET in the $^1\pi\pi^*$ excited state of the Ac···$H_2O$ complex is merely 0.40 eV, which is about twice the zero-point energy of the PT coordinate.[44] Such barriers can efficiently be tunnelled by H-atoms within the lifetime of the $^1\pi\pi^*$ state. It has recently been shown that the incorporation of Ac in carbon dots indeed leads to a strong increase of photocatalytic hydrogen evolution (methanol was used as sacrificial reagent).[45]

## 4. PHOTOOXIDATION OF WATER WITH THE HEPTAZINE CHROMOPHORE

There exists a huge number of N-heterocyclic dyes with absorption bands in the visible range of the spectrum which could be scrutinized as potential chromophores for the photooxidation of water as described by Eqs. (1-3). However, organic dyes typically are not photostable under irradiation with visible or UV light. N-heterocycles, in particular, generally possess low-lying $^3\pi\pi^*$ excited states below the lowest singlet excited state. Transfer of population from the singlet excited states to the lower-lying triplet states by intersystem crossing (ISC) typically initiates destructive photochemistry such as isomerization or fragmentation. Moreover, the population in long-lived triplet states may activate atmospheric triplet oxygen to highly reactive singlet oxygen which readily destroys organic materials. The activation of oxygen by long-lived triplet states is a long-standing problem in photocatalytic water splitting with TM complexes.[46, 47]

As shown by Antonietti, Maeda and coworkers, g-$C_3N_4$, consisting of Hz building blocks linked by imine groups, exhibits photocatalytic activity for hydrogen evolution with near-UV light in the presence of sacrificial electron donors and also exhibits high photostability under long-term irradiation.[13] As a purified molecule, heptazine (Hz) was first synthesized and structurally characterized by Leonard and coworkers in 1982[48] and the photoelectron spectrum was recorded.[49] Because the Hz molecule hydrolyses rapidly in the presence of traces of water, it attracted little interest in synthetic organic chemistry and only few experimental data on the photophysics and photochemistry of the Hz molecule are available. Two derivatives of Hz were synthesized and spectroscopically characterized by the Adachi group and were shown to be promising chromophores for organic light-emitting diodes (OLEDs).[50, 51] Recently. Schlenker and coworkers synthesized another Hz derivative, trianisole-heptazine (TAHz, see Fig. 1e), and showed that this molecular chromophore is, in contrast to Hz itself, chemically as well as photochemically highly stable.[21] Insights into the mechanisms of the photoreactivity of TAHz with water molecules were obtained by time-resolved luminescence measurements.[21]

The exceptional photostability and photocatalytic activity of Hz-based polymers as well as molecular species derived from Hz suggest that this chromophore may have unique spectroscopic and photochemical properties among N-heterocycles. Recent *ab initio* electronic-structure calculations for the excited states of Hz, TAHz and several other derivatives indeed revealed that the Hz chromophore exhibits highly unusual features which distinguish this chromophore from typical N-heterocycles.[22] Besides the very bright $^1\pi\pi^*$ state (calculated at 4.43 eV in Hz and 3.62 eV in TAHz with oscillator strengths of 0. 54 and 2.38, respectively) Hz chromophores possess a characteristic low-lying ($\approx$ 2.6 eV) $^1\pi\pi^*$ state ($S_1$) which is dark in absorption and emission, see Table 1. The transition dipole moment of the $S_1$ state with the electronic ground state is suppressed by the specific non-overlapping structures of HOMO and LUMO and additionally by symmetry selection rules in trigonal symmetry.[22] A low-lying dark $^1\pi\pi^*$ state is so unusual in N-heterocycles that the $S_1(\pi\pi^*)$ state of Hz derivatives occasionally has incorrectly been assigned as $^1n\pi^*$ state.[52] Moreover, the $S_1$ and $T_1$ states of Hz and its derivatives exhibit the very rare phenomenon of singlet-triplet inversion, that is, the vertical (and adiabatic) excitation energy of the $T_1$ state is higher than the vertical (and adiabatic) excitation energy of the $S_1$ state,[53-55] see Table 1. This unusual ordering of $S_1$ and $T_1$ states eliminates the decay channel of ISC for the $S_1$ state (in addition, $S_1$-$T_1$ SO coupling is forbidden by symmetry in first order in Hz and TAHz[55]). Together, these features of the $S_1$ state result in an exceptionally long lifetime of the $S_1$ state (measured as 287 ns for TAHz in toluene) and the remarkable fact that TAHz is strongly luminescent (quantum yield of 0.7 in toluene) despite the symmetry-forbidden transition dipole moment.[21]

The exceptionally long-lived $S_1$ state of Hz chromophores is of great significance for photocatalysis. The $S_1$ state plays the role of a reservoir state which allows the storage of a substantial part of the photon energy for timescales which are sufficient for productive photochemical transformations. The absorption of a photon by the strongly allowed second $^1\pi\pi^*$ state is followed by radiationless relaxation to the dark $S_1$ state on sub-picosecond timescales. This constellation resolves the problem that a large absorption cross section of a given state necessarily implies a short fluorescence lifetime. In addition, the singlet-triplet inversion in Hz chromophores implies the absence of a long-lived population in the lowest triplet state. The singlet-triplet inversion in Hz-based materials likely is one of the reasons for the exceptional photostability of these materials.

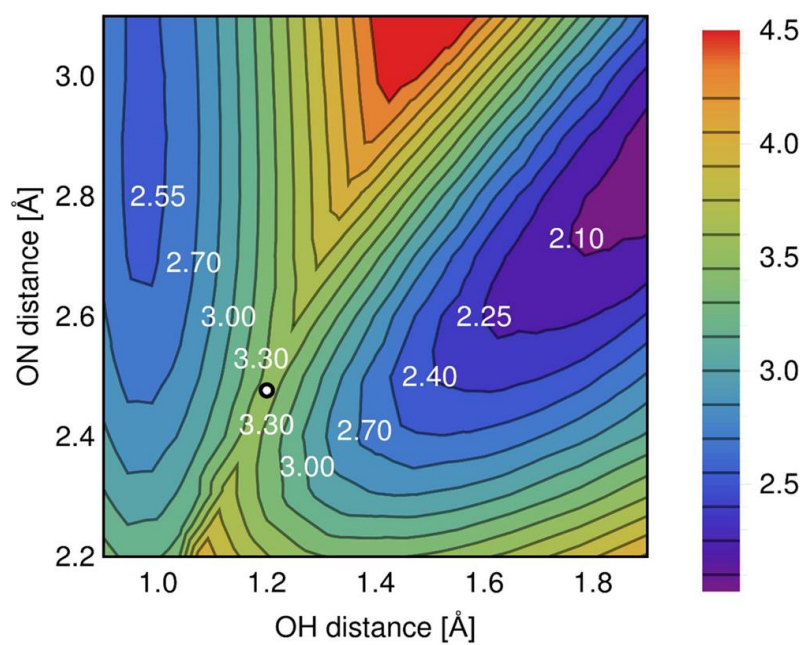

*Fig. 2. Relaxed PE surface of the lowest excited state of the Hz···H₂O complex, calculated with the ADC(2)/cc-pVDZ method. For fixed OH and NH distance, the energy of the $S_1$ state was minimized with respect to all other nuclear degrees of freedom. The circle indicates the saddle point which separates the Franck-Condon minimum of the locally excited $S_1(\pi\pi^*)$ state of Hz from the energy minimum of the HzH···OH biradical. Reprinted from ref. 22. Copyright 2017 American Chemical Society.*

The mechanism of the excited-state PCET reaction in the Hz···H₂O hydrogen-bonded complex can be visualized by the two-dimensional PE surface of the lowest excited singlet state of this complex, which is displayed in Fig. 2. The OH bond length of the hydrogen-bonded H₂O molecule represents the PT reaction coordinate. The second coordinate in Fig. 2 is the distance of the H-atom-donating oxygen atom of the water molecule from the H-atom-accepting nitrogen atom of Hz. The energy surface in Fig. 2 is relaxed with respect to all other nuclear coordinates of the complex. The well on the left hand side in Fig. 2 represents the locally excited $S_1$ state of Hz. The deep well on the right hand side in Fig. 2 represents the HzH···OH biradical. The barrier separating the two wells is 0.75 eV according to this calculation[22]. H-atom tunnelling through this barrier in the TAHz···$H_2O$ complex is responsible for the shortening of the $S_1$ lifetime from 287 ns in toluene to 16 ns in water.[21]

While characterizations of the excited-state PE landscape can provide a qualitative sketch of the photochemistry, a more detailed mechanistic understanding requires explicit computational simulations of the nonadiabatic nuclear dynamics. For the Hz···$H_2O$ complex, first-principles quasi-classical surface-hopping simulations were performed by Ullah et al.[56] at the level of time-dependent density function theory (TDDFT) and by Huang and Domcke[24] with the ADC(2) (algebraic-diagrammatic construction of second order) wave-function-based *ab initio* method. In the work of ref.[24], the absorption spectrum of the Hz···$H_2O$ complex was first calculated with the nuclear ensemble method[57] by sampling the positions and momenta of the nuclei from the Wigner distribution[58] in the electronic ground state and calculation of the vertical excitation energies and oscillator strengths of the excited states. The initial conditions for the sampling the photochemical dynamics were selected from a narrow slice of energies around the maximum of the absorption band, which corresponds to the vertical excitation of the bright $^1\pi\pi^*$ state of Hz. The forces for the propagation of the trajectories were computed "on the fly" with the ADC(2) method. Surface-hopping transitions are stochastically computed with probabilities determined with the quasiclassical Landau-Zener formula at avoided crossings of pairs of adiabatic PEs.[59] From a sufficiently large swarm of trajectories (typically ≈ 100), the time-dependent population probabilities of electronic states and the nuclear dynamics of the complex were evaluated. While the approximation of an aqueous environment by a single water molecule is a drastic simplification, studies of the photoinduced dynamics for smaller chromophores in larger water clusters have shown that the qualitative photochemical phenomena are clearly expressed already in clusters with a single water molecule.[41, 42] More details on the methodology and results of these ab initio nonadiabatic dynamics simulations can be found in ref.[24]

A schematic illustration of the time-resolved electronic population flow in the photoinduced dynamics of the Hz···$H_2O$ complex up to 100 fs is given by Fig. 3. For clarity, the higher excited electronic states (two bright $^1\pi\pi^*$ states and three $^1n\pi^*$ states) are collectively represented as "high LE" (locally excited) states in Fig. 3. In addition to the locally excited states, there exists a charge-transfer (CT) state which arises from the transfer of the hole in the HOMO of Hz to the water molecule. In the time window of 100 fs, the majority of the high LE states (62%) decays directly via internal conversion to the $S_1$ state. A fraction of 38% of the initial excited-state population is transferred to the reactive CT state. Part of this population (29%) flows to the $S_1$ state at the conical intersection of the CT state with the $S_1$ state, while 9% of trajectories remain in the CT state and hit the conical intersection of the CT state with the $S_0$ state. 91% of the initial excited electronic population is thus collected in the $S_1$ state, confirming it reservoir character. The branching ratio at the CT-$S_0$ conical intersection could not be determined in this simulation because the single-reference ADC(2) method is not reliable in the vicinity of this intersection. It should be noted that the branching ratios in Fig. 3 only represent the ultrafast (< 100 fs) component of the electronic population dynamics. The population in the $S_1$ reservoir state can yield HzH and OH radicals on much longer timescales by H-atom tunnelling through the barrier which separates the minimum of the $S_1$ PE surface from the biradical channel (see Fig. 2). The timescale of this tunnelling process is expected to be in the picosecond to nanosecond range which is beyond the range which can be covered with ab initio on-the-fly dynamics simulations.

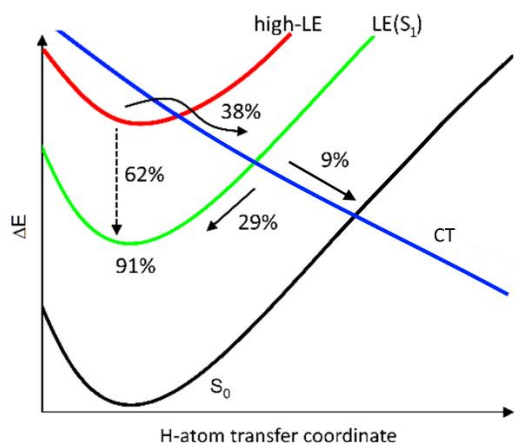

*Fig. 3. Schematic overview of the photochemical dynamics of the Hz-$H_2O$ complex within 100 fs. The numbers give the branching ratios of the nonadiabatic processes. Reprinted from ref. 24. Copyright 2021 American Chemical Society.*

The excitation energy of the bright $^1\pi\pi^*$ state (the "optical band gap") can be tuned by substituents on the Hz core, TAHz being an example. This is illustrated by Table 2, which shows the excitation energies of the $^1\pi\pi^*$ states and the CT state in the complexes of Hz, TAHz, trichloroheptazine (TClHz), and tricyanoheptazine (TCNHz) with a water molecule (the $^1n\pi^*$ states are omitted for clarity in Table 2). While the excitation energy of the $S_1(\pi\pi^*)$ state is little affected by substituents, the energy of the bright $^1\pi\pi^*$ state is stabilized by about 0.9 eV in TAHz relative to Hz. The vertical excitation energy of the CT state, on the other hand, is lowered by electron withdrawing groups such as CN, while the methoxy substituent raises the energy of the CT state substantially. The energy of the CT state varies from 6.15 eV in the TAHz⋯$H_2O$ complex to 4.13 eV in the TCNHz⋯$H_2O$ complex. As is schematically illustrated in Fig. 4, the vertical shift of the PE function of the CT state modulates the PCET barrier which results from the crossing of the PE function of the CT state with the PE function of the $S_1$ state. Quantitatively, the height of the PCET barrier on the $S_1$ potential-energy surface decreases from 1.08 eV for TAHz⋯$H_2O$ to 0.59 eV for TCNHz⋯$H_2O$.[23] Since the H-atom tunnelling rate depends exponentially on the barrier height, these data imply that the PCET rate can be substantially tuned by functionalization of the Hz core.

Table 2. Vertical excitation energies (in eV) and oscillator strengths (in parentheses) of the lowest three locally-excited $^1\pi\pi^*$ states and the lowest charge-transfer state of hydrogen-bonded complexes of Hz and three Hz derivatives with a water molecule, calculated with the ADC(2)/cc-pVDZ method at MP2-optimized ground-state equilibrium geometries. Oscillator strengths are given in parentheses. The charge-transfer state is marked in bold face.

| Hz···$H_2O$ | TClHz···$H_2O$ | TCNHz···$H_2O$ | TAHz···$H_2O$ |
|---|---|---|---|
| $S_1(\pi\pi^*)$ 2.59 (0.0000) | $S_1(\pi\pi^*)$ 2.79 (0.0001) | $S_1(\pi\pi^*)$ 2.33 (0.0001) | $S_1(\pi\pi^*)$ 2.65 (0.0013) |
| $S_5(\pi\pi^*)$ 4.43 (0.2578) | $S_5(\pi\pi^*)$ 4.48 (0.3250) | **$S_5$(CT) 4.13 (0.0001)** | $S_2(\pi\pi^*)$ 3.50 (1.0829) |
| $S_6(\pi\pi^*)$ 4.43 (0.2832) | $S_6(\pi\pi^*)$ 4.48 (0.2964) | $S_6(\pi\pi^*)$ 4.16 (0.3111) | $S_3(\pi\pi^*)$ 3.59 (1.2080) |
| **$S_{13}$(CT) 5.40 (0.0019)** | **$S_8$(CT) 4.96 (0.0024)** | $S_7(\pi\pi^*)$ 4.17 (0.2865) | **$S_n$(CT) 6.15 (0.0369)** |

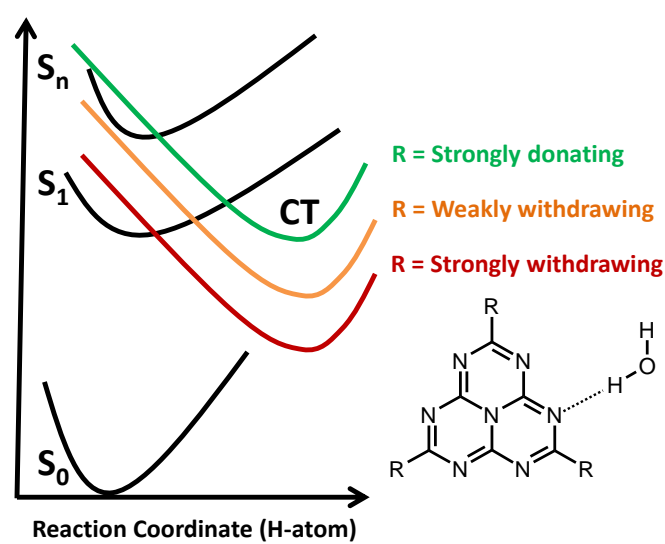

*Fig. 4. Schematic energy landscape diagram for substituted Hz···$H_2O$ complexes, illustrating the relationship between the vertical excitation energy of the CT state and the PCET barrier on the PE surface of the $S_1$ state. Increasing electron-withdrawing character of the substituent R lowers the energy of the CT state and the PCET barrier in the $S_1$ state. Adapted from ref. 23. Copyright 2020 American Chemical Society.*

On the experimental side, Schlenker and coworkers devised a model system for more detailed investigations of the PCET reactivity of Hz chromophores by considering the complexation of TAHz with functionalized phenols in toluene solution.[60, 61] Phenol (H-PhOH) is a sacrificial electron donor, which is reflected by the fact that the vertical excitation energy of the CT state (corresponding to hole transfer from H-PhOH to TAHz) in the TAHz-H-PhOH complex is 2.7 eV lower in energy that the corresponding CT state in the TAHz-$H_2O$ complex. Considering a series of para-substituted phenol substrates, R-PhOH, the hydrogen-bond strength with TAHz and the redox potential were systematically tuned. Increase of the electron-donating strength of R lowers the oxidation potential of R-PhOH and thereby lowers the energy of the CT state of the complex, which in turn lowers the barrier of the PCET reaction on the $S_1$ potential-energy surface, as schematically illustrated in Fig. 5.

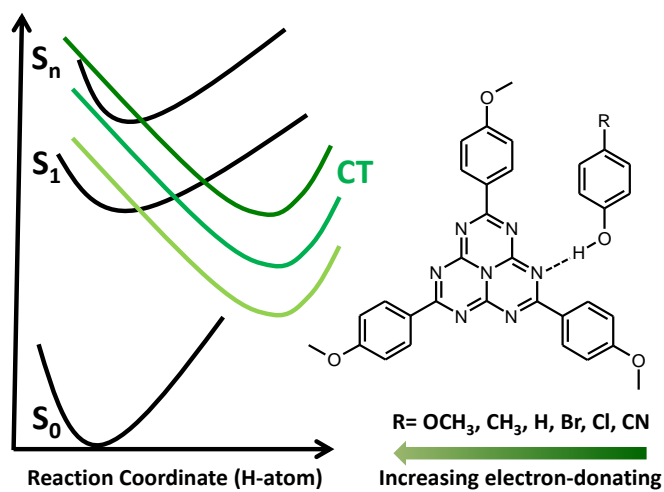

*Fig. 5. Schematic drawing of the PE functions of the locally excited states of TAHz ($S_1$, $S_n$) and of the CT state (from R-PhOH to TAHz) for different substituents R on the phenol substrate. Increasing electron donating strength of R lowers the energy of the CT state and the PCET barrier in the $S_1$ state. Adapted from ref. 60. Copyright 2019 American Chemical Society.*

Calculations of relaxed 2D energy surfaces analogous to Fig. 3 for complexes of TAHz with H-PhOH, CN-PhOH, and MeO-PhOH predict that the PCET barrier varies from 0.37 eV for CN-PhOH to zero (barrierless) for MeO-PhOH. These computational findings compare well with the measured luminescence quenching rates of the $S_1$ state.[60] Building on this rather detailed understanding of the photochemistry of TAHz-R-PhOH complexes, Schlenker and coworkers were able to demonstrate the possibility of active control of excited-state PCET reactivity in TAHz-R-PhOH by pump-push-probe experiments.[62]

## 5. WATER PHOTOOXIDATION WITH CARBON NITRIDES AND CONJUGATED POLYMERS

In the extensive literature on hydrogen evolution with PCNs, water oxidation and hydrogen evolution (usually with sacrificial electron donors, such as triethylamine (TEA), triethanolamine (TEOA) or methanol (MeOH), are interpreted in terms of the indirect mechanism established for photocatalytic water splitting with metal-oxide semiconductors, as discussed above. However, excitons in polymeric or crystalline carbon nitrides are Frenkel excitons. The binding energies of these molecular excitons are of the order of several electron volts, whereas the binding energies of Mott-Wannier excitons in metal-oxide semiconductors typically are of the order of 10 meV.[4] In contrast to the extensive data on carrier dynamics in metal-oxide semiconductors, quantitative data on transport coefficients are lacking for g-$C_3N_4$ and related polymeric organic materials.[20] Excited-state lifetime data have been obtained for carbon nitrides with femtosecond to nanosecond transient absorption spectroscopy,[63-67] but these data can alternatively be interpreted in a localized molecular picture.[63] No data seem to exist on the rates and yields of redox reactions with water or sacrificial reagents at the polymer-solvent interface.

$H_2O$ does not adsorb dissociatively on organic polymers, in contrast to $TiO_2$. Considering that the ionization potential of Hz is 9.6 eV[49] and the ionization potential of $H_2O$ is 12.6 eV in the gas phase[68] and 11.2 eV in a liquid jet,[69] it is unlikely that a relaxed hole in a carbon nitride material can oxidize water. Overall, there is no evidence supported by quantitative experimental data that the indirect water-splitting paradigm established for inorganic semiconductors can be transferred one-to-one to water oxidation and hydrogen evolution with organic polymers.

On the theoretical side, band structures of periodic models of $g-C_3N_4$ have been computed with DFT.[70-72] Simulations of the dynamics of photoinduced reaction steps were performed at the DFT level for extended models of $g-C_3N_4$ with periodic boundary conditions by Ma and coworkers[73, 74] and Meng and coworkers.[75] In both simulations, the water oxidation step occurs locally on a Hz building block by H-atom transfer from a hydrogen-bonded water molecule to a peripheral N-atom of a Hz unit as found in the molecular simulations discussed above. Ma and coworkers additionally discussed follow-up dark reactions of the nascent OH radical.[74] Interestingly, Muuronen et al. found evidence for a similar spatially localized water oxidation mechanism in nonadiabatic dynamics simulations of photoinduced water oxidation on small $TiO_2$ nanoparticles.[76]

An alternative extensively investigated class of metal-free organic materials for hydrogen evolution photocatalysis are conjugated linear polymers. Already in the early 1990s, Yanagida and coworkers reported hydrogen generation with poly(p-phenylene) and poly(pyridine-diyl) polymers in the presence of a dispersed noble-metal co-catalyst and a sacrificial electron donor.[77, 78] More recently, the hydrogen evolution reaction with organic polymer photocatalysts was systematically explored by Cooper and coworkers.[20, 79, 80] Like in the polymeric carbon nitride materials, the photocatalytic mechanisms in these materials are currently poorly understood. Prentice and Zwijnenburg recently discussed the pros and cons of the charge-carrier mechanism and the radical-pair mechanism in conjugated polymers.[81, 82] Evidence has been reported that conjugated polymers containing N-heterocycles are superior photocatalysts and that the hydrogen evolution yield seems to increase with the nitrogen content.[83-85] Moreover, it has been observed that small oligomers seem to be more efficient photocatalysts than extended polymers of the same material.[86] These findings are hints that photooxidation of the substrate at local nitrogen centers may play a role in the photocatalysis. The observations are more easily explained in the radical-pair mechanisms than in the charge-carrier mechanism.

## 6. HARVESTING OF HYDROGEN

To close the photocatalytic cycle after radical pair formation in a chromophore-water complex, the excess hydrogen atom has to be detached or abstracted from the reduced chromophore. Several reaction mechanisms are conceivable, both photoinduced (excited-state) reactions as well as thermal (ground-state) reactions. The most direct recovery of the chromophore is by the photodetachment of the excess H-atom from the AH radical

$$AH + h\nu \rightarrow A + H. \quad (4)$$

Calculations for the PyH, AcH, HzH and related radicals have shown that these species possess low-lying moderately absorbing $^2\pi\pi^*$ states as well as a low-lying dark $^2\pi\sigma^*$ state.[22, 39, 44] While the $^2\pi\pi^*$ states are bound with respect to the NH bond distance, the potential-energy function of the $^2\pi\sigma^*$ state is dissociative. When the $^2\pi\sigma^*$ state is populated by vibronic intensity borrowing or by radiationless relaxation from higher $^2\pi\pi^*$ states, it can drive an ultrafast (non-statistical) photodetachment reaction, resulting in the recovery of the chromophore and a free hydrogen atom. For the PyH radical, the photodetachment reaction was detected in a molecular beam experiment[38] and the reaction dynamics was studied with nonadiabatic time-dependent wave-packet calculations.[87]

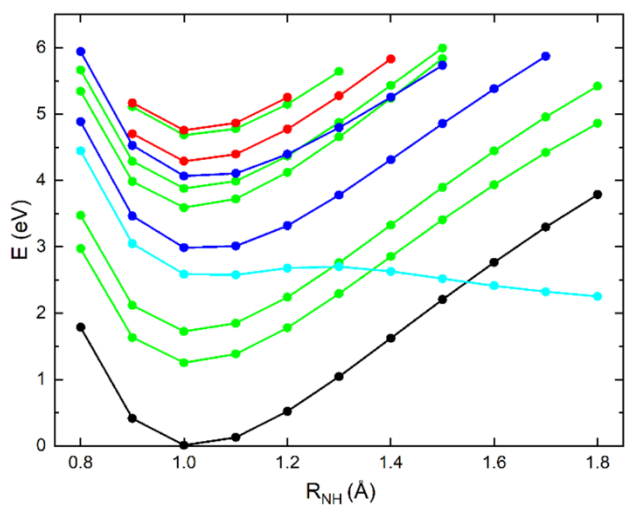

*Fig. 6. PE functions of planar $HzH_2$ for a rigid scan along the NH dissociation coordinate $R_{NH}$ at the ADC(2)/cc-pVDZ level. Black: ground state, green: $\pi\pi^*$ states, red: $n\pi^*$ states, light blue and dark blue: $\pi\sigma^*$ states.*

Hz, TAHz and related photooxidants possess several peripheral nitrogen atoms. These radicals can disproportionate according to

$$AH + AH \rightarrow AH_2 + A. \quad (5)$$

Calculations predict that the disproportionation reaction is exothermic by 1.5 eV for HzH.[24] The doubly reduced $AH_2$ molecule is a stable closed-shell species and therefore suitable for long-term chemical storage of hydrogen atoms. The hydrogen atoms can be recovered by photodissociation of $AH_2$

$$AH_2 + h\nu \rightarrow AH + H. \quad (6)$$

The AH radicals formed by the reaction (6) will disproportionate again such that eventually A is regenerated and H radicals are produced and can be used for reduction reactions.

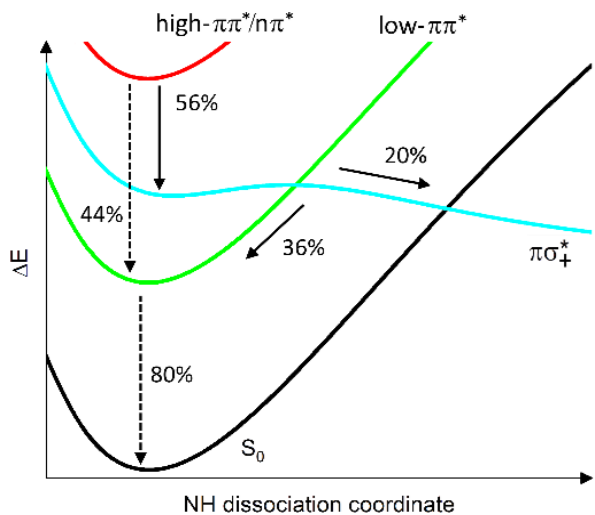

*Fig. 7. Schematic overview of the photochemical dynamics of HzH$_2$ within 150 fs. The numbers give the branching ratios of nonadiabatic processes. 20% of the trajectories encounter CIs of the diabatic πσ\* state with the electronic ground state at stretched NH bond lengths. Reprinted from ref. 24. Copyright 2021 American Chemical Society.*

The mechanism of the photodetachment reaction from HzH$_2$ is illustrated by Fig. 6. HzH$_2$ possesses two low-lying $^1$πσ\* states, one for each NH group. When one of the two N-H bond lengths is stretched, the PE function of the corresponding $^1$πσ\* state develops a dissociative character, see the light blue curve in Fig. 6. The dissociative $^1$πσ\* state can be populated either directly via a vibronically induced transition or indirectly by excitation of higher $^1$ππ\* state followed by rapid internal conversion to the $^1$πσ\* state. Fig. 6 illustrates that three conical intersections (with two low-lying $^1$ππ\* state and with the S$_0$ state) have to be passed diabatically in the H-atom detachment reaction. The nonadiabatic dynamics of this photodissociation process was investigated with *ab initio* surface-hopping trajectory simulations analogous to those for the Hz⋯H$_2$O complex discussed above, see ref.[24] for details. To illustrate the essential aspects of the reaction dynamics, the electronic states of HzH$_2$ are grouped as ground state (S$_0$), low $^1$ππ\* states (S$_1$, S$_2$), $^1$πσ\* state (S$_3$), and high-lying $^1$ππ\*/$^1$nπ\* states (S$_n$, n > 3). Fig. 7 illustrates the electronic population flow, assuming excitation of the $^1$πσ\* PE surface from one of the higher $^1$ππ\* states. A fraction of 56% of the initial excited-state population is transferred to the dissociative $^1$πσ\* state, while 44% of the population relaxes directly to the low-lying $^1$ππ\* states. At the intersection of the $^1$πσ\* PE surface with the PE surfaces of the low-lying $^1$ππ\* states, 36% of the population is captured in the low-lying $^1$ππ\* states, while 20% pass to the conical intersections of the $^1$πσ\* energy with the S$_0$ energy. Remarkably, no long-lived excited-state population exists in HzH$_2$. The dynamics is essentially completed within 150 fs. This feature is in stark contrast to the dynamics of the Hz⋯H$_2$O complex (Fig. 3) in which most of the photoexcited population is trapped in the S$_1$ state within 100 fs.

Alternatively to the light-driven hydrogen harvesting reactions of Eqs. (4) and (6), the chromophore A and molecular hydrogen can be recovered by a thermal radical-radical recombination reaction

$$AH + AH \rightarrow 2A + H_2. \tag{7}$$

The reaction (7) is exothermic if the dissociation energy of the AH radical is less than half of the dissociation energy (4.48 eV) of $H_2$. According to calculations, this condition is fulfilled for most of the AH radicals mentioned above. After disproportionation of AH (Eq. (5)), $H_2$ also may be recovered by a thermal decomposition reaction with a noble-metal catalyst

$$AH_2 \rightarrow A + H_2. \tag{8}$$

## 7. MANAGING THE PHOTOGENERATED OH RADICALS

There is no free lunch in water splitting photocatalysis. The long-standing carrier recombination problem in the indirect water oxidation mechanism is mirrored by the radical recombination problem in the direct water oxidation mechanism discussed herein. OH radicals are highly mobile and extremely reactive. Clearly, recombination of the photogenerated OH and H radicals is unproductive and has to be avoided. In addition, recombination of the photogenerated OH radicals with the reduced chromophore (AH) is undesirable, since this will destroy the precious photocatalyst. The thermochemistry of the latter reaction has been studied for the example of Hz and several derivatives of Hz.[23] The structures and relative energies of the Hz$\cdots$H$_2$O complex, the HzH$\cdots$OH biradical, and possible products of the radical recombination reaction are displayed in Fig. 8a-d. The OH radical may attach to one of the CH groups of Hz via a covalent bond, which results in a closed-shell electronic structure (Fig. 8c). This photohydrate of Hz is merely 1.06 kcal/mol higher in energy than the hydrogen-bonded Hz$\cdots$H$_2$O complex at the MP2/cc-pVDZ level. The breaking of the C-N covalent bond requires about 12 kcal/mol. From this structure, a downhill reaction leads to the open amino-aldehyde form (Fig. 8d). According to these results, the open form of the photohydrate of Hz is thermochemically stable. In contrast to Hz, the photohydratre of Hz does not absorb in the visible range. The photoinduced conversion of Hz to the photohydrate is thus an irreversible process, which explains the apparently spontaneous hydrolysis of the Hz molecule.[48] For TAHz, on the other hand, the calculations predict that the photohydrate is metastable by about 6 kcal/mol.[23] The photohydrate of TAHz will thus convert back to TAHz and $H_2O$ by a thermal reaction in the dark, which implies that TAHz is a self-healing water-oxidation photocatalyst. It is likely that two properties of TAHz, the inversion of singlet and triplet states (avoiding the activation of oxygen) and the capacity of self-healing, are responsible for its exceptional photostability. While self-healing after photohydration likely is an essential feature of any practically viable water-splitting photocatalyst, the energy of the photon nevertheless is lost when photohydration occurs.

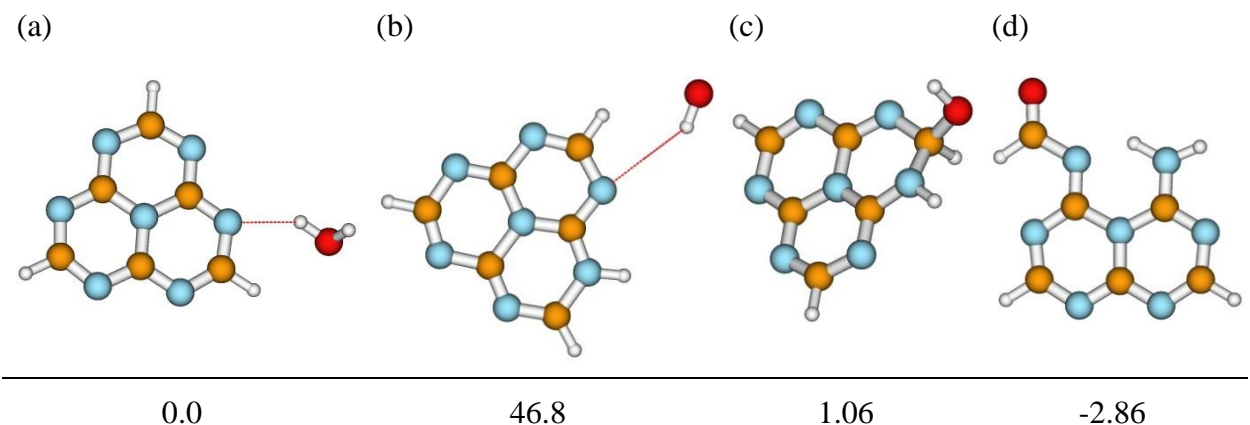

| (a) | (b) | (c) | (d) |
|---|---|---|---|
| 0.0 | 46.8 | 1.06 | -2.86 |

Fig. 8: Structures and relative energies (in kcal/mol) of the Hz···$H_2O$ complex (a), the HzH···OH biradical (b), and two photohydrates of Hz (c, d), computed at the MP2/cc-pVDZ level.

To avoid deleterious reactions of the OH radicals with the photocatalyst, efficient scavenging and disposal of OH radicals is essential. This problem seems to have found relatively little attention so far. Noting that OH radicals are central to any combustion process, we propose that OH radicals may be disposed of by low-temperature combustion of a suitable sacrificial reactant, for example methane. The oxidation of methane to methanol yields a valuable chemical feedstock and a liquid fuel. Methane is one of the few molecules with an ionization potential (12.7 eV) which is higher than the ionization potential of water (12.6 eV).[68] Moreover, methane does not form hydrogen bonds with N-heterocycles. The presence of methane therefore does not interfere with the oxidation of water on Hz-based photocatalysts. In addition, the bond energy of methane (4.41 eV or 425 kcal/mol) is significantly lower than the bond energy of water (5.11 eV or 493 kcal/mol). The oxidation of $CH_4$ to $CH_3OH$ by OH radicals is thus strongly exothermic. Two OH radicals are consumed by the formation of one $CH_3OH$ molecule.

Imagining an optically transparent porous sieve decorated with a Hz-based photocatalyst, the photocatalytic water-oxidation and methane-oxidation reactions could take place jointly in the gas phase. The partial pressure of $CH_4$ could be adjusted to minimize the recombination of OH radicals with the Hz-based photocatalyst which accumulates and stores hydrogen atoms from water. It has been shown that the reaction of $CH_4$ with OH radicals which were photocatalytically generated with a $TiO_2$-supported iron catalyst can yield methanol with a conversion rate of 90%.[88] When methane is recovered from biomass, this process yields green methanol and green $H_2$. Notably, the hydrogen atoms evolved in this process are 100% taken from water.

# 8. CONCLUSIONS

In the current literature, PCNs are perceived as semiconducting materials and the emphasis of the current research efforts is on the improvement of the efficiency of charge carrier generation and charge migration in these materials. The suppression of the intrinsically high fluorescence quantum yield of PCNs by structural modifications or doping is considered desirable, based on the assumption that detrimental trapping of free charge carriers at defects and charge recombination are the origin of the fluorescence. If, however, excitons in PCNs never dissociate and the oxidation of water (or sacrificial electron donors) occurs via an excited-state photochemical reaction, then a high fluorescence quantum yield is a desirable property indicating a long excited-state lifetime. These considerations underline that a full understanding of the precise mechanisms of the photoinduced transformations is essential for the rational development of substantially improved photocatalytic materials.

In this perspective, we put the spectroscopic and photochemical properties of molecular Hz into the spot light. Albeit Hz was synthesized and structurally characterized already in 1982,[48] the spectroscopic properties were never explored, since Hz hydrolyses quickly and irreversibly. Only recently, *ab initio* calculations revealed the outstanding properties of the Hz chromophore for photocatalysis. Hz possesses a strongly absorbing second $^1\pi\pi^*$ state, the excitation energy of which is widely tuneable by substituents. It further possesses a low-lying and exceptionally long-lived $S_1(\pi\pi^*)$ state which is located below the lowest triplet state. These features allow the storage of electronic excitation energy in the $S_1$ state of the Hz chromophore for hundreds of nanoseconds, such that the excitation energy can be exploited for diffusion-controlled photochemical transformations. The absence of a long-lived triplet state eliminates the activation of atmospheric oxygen to deleterious singlet oxygen. Moreover, the self-healing capacity of certain derivatives of Hz after oxidation by OH radicals contributes to the exceptional photostability of Hz-based materials. The recent synthesis and spectroscopic characterization of TAHz enabled an unprecedentedly detailed comparison of the computational predictions with experiment.

Ideally, the photocatalytic properties of organic materials should be optimized in two (or more) steps. In the first step, the properties should be optimized at the molecular level, that is, for individual chromophores, by judicious comparisons of experiment data with high-level *ab initio* theory data. In a second step, structurally controlled materials (such as, e.g., covalent organic frameworks or molecular sieves) should be assembled from the pre-optimized building blocks and the properties of the assemblies should be optimized in a systematic manner with support from condensed-phase theory. New levels of functionality of photocatalytic materials may thus be achieved in the future by synergistic efforts of synthesis, spectroscopy, photochemistry, and theory.

# REFERENCES

1. Fujishima, A.; Honda, K. Electrochemical Photolysis of Water at a Semiconductor Electrode. *Nature* **1972,** *238*, 37-38.
2. Cowan, A. J.; Durrant, J. R. Long-Lived Charge Separated States in Nanostructured Semiconductor Photoelectrodes for the Production of Solar Fuels. *Chem. Soc. Rev.* **2013,** *42*, 2281-2293.
3. Schneider, J.; Matsuoka, S.; Takeuchi, M.; Zhang, J.; Horiuchi, Y.; Anpo, M.; Bahnemann, D. W. Understanding $TiO_2$ Photocatalysis: Mechanisms and Materials. *Chem. Rev.* **2014,** *114*, 9919-9986.
4. Takanabe, K. Photocatalytic Water Splitting: Quantitative Approaches Toward Photocatalyst by Design. *ACS Catal.* **2017,** *7*, 8006-8022.
5. Wang, Q.; Domen, K. Particulate Photocatalysts for Light-Driven Water Splitting: Mechanisms, Challenges and Design Strategies. *Chem. Rev.* **2019,** *120*, 919-985.
6. Chen, X.; Shen, S.; Guo, L.; Mao, S. S. Semiconductor-Based Photocatalytic Hydrogen Generation. *Chem. Rev.* **2010,** *110*, 6503-6570.
7. Ager, J. W.; Shaner, M. R.; Walczak, K. A.; Sharp, I. D.; Ardo, S. Experimental Demonstrations of Spontaneous, Solar-Driven Photoelectrochemical Water Splitting. *Energy Env. Sci.* **2015,** *8*, 2811-2824.
8. Kang, D.; Kim, T. W.; Kubota, S. R.; Cardiel, A. C.; Cha, H. G.; Choi, K.-S. Electrochemical Synthesis of Photoelectrodes and Catalysts for Use in Solar Water Splitting. *Chem. Rev.* **2015,** *115*, 12839-12887.
9. Vanka, S.; Sun, K.; Zeng, G.; Pham, T. A.; Toma, F. M.; Ogitsu, T.; Mi, Z. Long-Term Stability Studies of a Semiconductor Photoelectrode in Three-Electrode Configuration. *J. Mater. Chem. A* **2019,** *7*, 27612-27619.
10. Kistler, T. A.; Zeng, G.; Young, J. L.; Weng, L.-C.; Aldrige, C.; Wyatt, K.; Steiner, M. A.; Solarzano, J., O.; Houle, F. A.; Toma, F. M., et al. Emergent Degradation Phenomena Demonstrated on Resilient, Flexible, and Durable Integrated Photoelectrochemical Cells. *Adv. Energy Mater.* **2020**, No. 2002706.
11. Du, P.; Eisenberg, R. Catalysts Made of Earth-Abundant Elements (Co, Ni, Fe) for Water Splitting: Recent Progress and Future Challenges. *Energy Env. Sci.* **2012,** *5*, 6012-6021.
12. Blakemore, J. D.; Crabtree, R. H.; Brudvig, G. W. Molecular Catalysts for Water Oxidation. *Chem. Rev.* **2015,** *115*, 12974-13005.
13. Wang, X.; Maeda, K.; Thomas, A.; Takanabe, K.; Xin, G.; Carlsson, J. M.; Domen, K.; Antonietti, M. A Metal-Free Polymeric Photocatalyst For Hydrogen Production from Water under Visible Light. *Nat. Mater.* **2009,** *8*, 76-80.
14. Liebig, J. Über einige Stickstoff Verbindungen. *Ann. Pharm.* **1834,** *10*, 10.
15. Wang, Y.; Wang, X.; Antonietti, M. Polymeric Graphitic Carbon Nitride as a Heterogeneous Organocatalyst: From Photochemistry to Multipurpose Catalysis to Sustainable Chemistry. *Angew. Chem., Int. Ed.* **2012,** *51*, 68-89.
16. Schwarzer, A.; Saplinova, T.; Kroke, E. Tri-s-Triazines (s-Heptazines) - From a "Mystery Molecule" to Industrially Relevant Carbon Nitride Materials. *Coord. Chem. Rev.* **2013,** *257*, 2032-2062.
17. Ong, W.-J.; Tan, L.-L.; Ng, Y. H.; Yong, S.-T.; Chai, S.-P. Graphitic Carbon Nitride (g-$C_3N_4$)-Based Photocatalysts for Artificial Photosynthesis and Environmental Remediation: Are we a Step Closer to Achieving Sustainability? *Chem. Rev.* **2016,** *116*, 7159-7329.
18. Wen, J.; Xie, J.; Chen, X.; Li, X. A Review on g-$C_3N_4$-Based Photocatalysts. *Appl. Surf. Sci.* **2017,** *391*, 72-123.
19. Lau, V. W.; Lotsch, B. V. A Tour-Guide Through Carbon-Nitride-Land: Structure- and Dimensionality-Dependent Properties for Photo(Electro) Chemical Energy Conversion and Storage. *Adv. Energy Mater.* **2021**, No. 2101078.

20. Wang, Y.; Vogel, A.; Sachs, M.; Sprick, R. S.; Wilbraham, L.; Moniz, S. J. A.; Godin, R.; Zwijnenburg, M. A.; Durrant, J. R.; Cooper, A. I., et al. Current Understanding and Challenges of Solar-Driven Hydrogen Generation Using Polymeric Photocatalysts. *Nature Energy* **2019,** *4*, No. 746.

21. Rabe, E. J.; Corp, K. L.; Sobolewski, A. L.; Domcke, W.; Schlenker, C. W. Proton-Coupled Electron Transfer from Water to a Model Heptazine-Based Molecular Photocatalyst. *J. Phys. Chem. Lett.* **2018,** *9*, 6257-6261.

22. Ehrmaier, J.; Karsili, T. N. V.; Sobolewski, A. L.; Domcke, W. Mechanism of Photocatalytic Water Splitting with Graphitic Carbon Nitride: Photochemistry of the Heptazine-Water Complex. *J. Phys. Chem. A* **2017,** *121*, 4754-4764.

23. Ehrmaier, J.; Huang, X.; Rabe, E. J.; Corp, K. L.; Schlenker, C. W.; Sobolewski, A. L.; Domcke, W. Molecular Design of Heptazine-Based Photocatalysts: Effect of Substituents on Photocatalytic Efficiency and Photostability. *J. Phys. Chem. A* **2020,** *124*, 3698-3710.

24. Huang, X.; Domcke, W. Ab initio Nonadiabatic Surface-Hopping Trajectory Simulations of Photocatalytic Water Oxidation and Hydrogen Evolution with the Heptazine Chromophore. *J. Phys. Chem. A* **2021,** *125*, 9917-9931.

25. Gerischer, H. A Mechanism of Electron Hole Pair Separation in Illuminated Semiconductor Particles. *J. Phys. Chem.* **1984,** *88*, 6096-6097.

26. Chen, M.; Straatsma, T. P.; Dixon, D. A. Molecular and Dissociative Adsorption of Water on $(TiO_2)_n$ Clusters, n = 1-4. *J. Phys. Chem. A* **2015,** *119*, 11406-11421.

27. Stecher, T.; Reuter, K.; Oberhofer, H. First-Principles Free-Energy Barriers for Photoelectrochemical Surface Reactions: Proton Abstraction at TiO$_2$(110). *Phys. Rev. Lett.* **2016,** *117*, 276001.

28. Yamakata, A.; Ishibashi, T.; Onishi, H. Water- and Oxygen-Induced Decay Kinetics of Photogenerated Electrons in TiO$_2$ and Pt/TiO$_2$: A Time-Resolved Infrared Absorption Study. *J. Phys. Chem. B* **2001,** *105*, 7258-7262.

29. Tang, J.; Durrant, J. R.; Klug, D. R. Mechanism of Photocatalytic Water Splitting in TiO$_2$. Reaction of Water with Photoholes, Importance of Charge Carrier Dynamics, and Evidence for Four-Hole Chemistry. *J. Am. Chem. Soc.* **2008,** *130*, 13885-13891.

30. Godin, R.; Kafizas, A.; Durrant, J. R. Electron Transfer Dynamics in Fuel Producing Photosystems. *Curr. Opin. Electrochem.* **2017,** *2*, 136-143.

31. Serpone, N.; Emeline, A. V.; Ryabchuk, V. K.; Kuznetsov, V. N.; Artemev, Y. M.; Horikoshi, S. Why do Hydrogen and Oxygen Yields from Semiconductor-Based Photocatalyzed Water Splitting Remain Disappointingly Low? Intrinsic and Extrinsic Factors Impacting Surface Redox Reactions. *ACS Energy Lett.* **2016,** *1*, 931-948.

32. Jacobsson, T. J. Photoelectrochemical Water Splitting: An Idea Heading Towards Obsolescence? *Energy Env. Sci.* **2018,** *11*, 1977-1979.

33. Fukuzumi, S. Development of Bioinspired Artificial Photosynthetic Systems. *Phys. Chem. Chem. Phys.* **2008,** *10*, 2283-2297.

34. Gust, D.; Moore, T. A.; Moore, A. L. Solar Fuels via Artificial Photosynthesis. *Acc. Chem. Res.* **2009,** *42*, 1890-1898.

35. Moore, G. F.; Brudvig, G. W. Energy Conversion in Photosynthesis: A Paradigm for Solar Fuel Production. *Annu. Rev. Condens. Matter Phys.* **2011,** *2*, 303-327.

36. Sobolewski, A. L.; Domcke, W. Computational Studies of Hydrogen-Bonded Molecular Systems. *J. Phys. Chem. A* **2007,** *111*, 11725-11735.

37. Weinberg, D. R.; Gagliardi, C. J.; Hull, J. F.; Murphy, C. F.; Kent, C. A.; Westlake, B. C.; Paul, A.; Ess, D. H.; McCafferty, D. G.; Meyer, T. J. Proton-Coupled Electron Transfer. *Chem. Rev.* **2012,** *112*, 4016-4093.

38. Esteves-Lopez, N.; Coussan, S.; Dedonder-Lardeux, C.; Jouvet, C. Photoinduced Water Splitting in Pyridine-Water Clusters. *Phys. Chem. Chem. Phys.* **2016,** *18*, 25637-25644.

39. Liu, X.; Sobolewski, A. L.; Borelli, R.; Domcke, W. Computational Investigation of the Photoinduced Homolytic Dissociation of Water in the Pyridine-Water Complex. *Phys. Chem. Chem. Phys.* **2013,** *15*, 5957-5966.

40. Liu, X.; Sobolewski, A. L.; Domcke, W. Photoinduced Oxidation of Water in the Pyridine-Water Complex: Comparison of the Singlet and Triplet Photochemistries. *J. Phys. Chem. A* **2014**, *118*, 7788-7795.
41. Pang, X.; Jiang, C.; Xie, W.; Domcke, W. Photoinduced electron-driven proton transfer from water to an N-heterocyclic chromophore: nonadiabatic dynamics studies for pyridine–water clusters. *Phys. Chem. Chem. Phys.* **2019**, *21*, 14073-14079.
42. Huang, X.; Aranguren, J.-P.; Ehrmaier, J.; Noble, J. A.; Xie, W.; Sobolewski, A. L.; Dedonder-Lardeux, C.; Jouvet, C.; Domcke, W. Photoinduced Water Oxidation in Pyrimidine-Water Clusters: A Combined Experimental and Theoretical Study. *Phys. Chem. Chem. Phys.* **2020**, *22*, 12502-12514.
43. Roy, S.; Ardo, S.; Furche, F. 5-Methoxyquinoline Photobasicity is Mediated by Water Oxidation. *J. Phys. Chem. A* **2019**, *123*, 6645-6651.
44. Liu, X.; Karsili, T. N. V.; Sobolewski, A. L.; Domcke, W. Photocatalytic Water Splitting with the Acridine Chromophore: A Computational Study. *J. Phys. Chem. B* **2015**, *119*, 10664-10672.
45. Fang, J.; Debnath, T.; Bhattacharyya, S.; Döblinger, M.; Feldmann, J.; Stolarczyk, J. K. Photobase Effect for Just-in-Time Delivery in Photocatalytic Hydrogen Generation. *Nature Commun.* **2020,** *11*, No. 5179.
46. Gust, D.; Moore, T. A.; Moore, A. L. Realizing Artificial Photosynthesis. *Faraday Discuss.* **2012,** *155*, 9-26.
47. Limburg, B.; Bouwman, E.; Bonnet, S. Molecular Water Oxidation Catalysts Based on Transition Metals and Their Decomposition Pathways. *Coord. Chem. Rev.* **2012,** *256*, 1451-1467.
48. Hosmane, R. S.; Rossman, M. A.; Leonard, N. J. Synthesis and Structure of Tri-s-triazine. *J. Am. Chem. Soc.* **1982,** *104*, 5497-5499.
49. Shabaz, M.; Urano, S.; LeBreton, P. R.; Rossman, M. A.; Hosmane, R. S.; Leonard, N. J. Tri-s-triazine: Synthesis, Chemical Behavior, and Spectroscopic and Theoretical Probes of Valence Orbital Structure. *J. Am. Chem. Soc.* **1984,** *106*, 2805-2811.
50. Li, J.; Nakagawa, T.; MacDonald, J.; Zhang, Q.; Nomura, H.; Miyazaki, H.; Adachi, C. Highly Efficient Organic Light-Emitting Diode Based on a Hidden Thermally Activated Delayed Fluorescence Channel in a Heptazine Derivative. *Adv. Mater.* **2013,** *25*, 3319-3323.
51. Li, J.; Nomura, H.; Miyazaki, H.; Adachi, C. Highly Efficient Exciplex Organic Light-Emitting Diodes Incorporating a Heptazine Derivative as an Electron Acceptor. *Chem. Commun.* **2014,** *50*, 6174-6176.
52. Li, J.; Zhang, Q.; Nomura, H.; Miyazaki, H.; Adachi, C. Thermally Activated Delayed Fluorescence from $^3n\pi^*$ to $^1n\pi^*$ Up-Conversion and its Application to Organic Light-Emitting Diodes. *Appl. Phys. Lett.* **2014,** *105*, No. 013301.
53. Ehrmaier, J.; Rabe, E. J.; Pristash, S. R.; Corp, K. L.; Schlenker, C. W.; Sobolewski, A. L.; Domcke, W. Singlet–Triplet Inversion in Heptazine and in Polymeric Carbon Nitrides. *J. Phys. Chem. A* **2019,** *123*, 8099-8108.
54. Sobolewski, A. L.; Domcke, W. Are Heptazine-Based Organic Light-Emitting Diode Chromophores Thermally Activated Delayed Fluorescence or Inverted Singlet-Triplet Systems? *J. Phys. Chem. Lett.* **2021,** *12*, 6852-6860.
55. Dinkelbach, F.; Bracker, F.; Kleinschmidt, M.; Marian, C. M. Large Inverted Singlet-Triplet Energy Gaps Are Not Always Favorable for Triplet Harvesting: Vibronic Coupling drives the (Reverse) Intersystem Crossing in Heptazine Derivatives. *J. Phys. Chem. A* **2021,** *125*, 10044-10051.
56. Ullah, N.; Chen, S.; Zhao, Y.; Zhang, R. Photoinduced Water-Heptazine Electron-Driven Proton Transfer: Perspective for Water Splitting with g-$C_3N_4$. *J. Phys. Chem. Lett.* **2019**, *10*, 4310-4316.
57. Crespo-Otero, R.; Barbatti, M. Spectrum Simulation and Decomposition with Nuclear Ensemble: Formal Derivation and Application to Benzene, Furan and 2-Phenylfuran. *Theor. Chem. Acc.* **2012,** *131*, No. 1237.
58. Hillery, M.; O'Connell, R. F.; Scully, M. O.; Wigner, E. P. Distribution Functions in Physics: Fundamentals. *Phys. Rep.* **1984,** *106*, 121-167.

59. Xie, W.; Sapunar, M.; Došlić, N.; Sala, M.; Domcke, W. Assessing the performance of trajectory surface hopping methods: Ultrafast internal conversion in pyrazine. *J. Chem. Phys.* **2019,** *150*, 154119.
60. Rabe, E. J.; Corp, K. L.; Huang, X.; Ehrmaier, J.; Flores, R. G.; Estes, S. L.; Sobolewski, A. L.; Domcke, W.; Schlenker, C. W. Barrierless Heptazine-Driven Excited-State Proton-Coupled Electron Transfer: Implications for Controlling Photochemistry on Carbon Nitrides and Aza-Arenes. *J. Phys. Chem. C* **2019,** *123*, 29580-29588.
61. Hwang, D.; Schlenker, C. W. Photochemistry of Carbon Nitrides and Heptazine Derivatives. *Chem. Commun.* **2021,** *57*, 9330-9353.
62. Corp, K. L.; Rabe, E. J.; Huang, X.; Ehrmaier, J.; Kaiser, M. E.; Sobolewski, A. L.; Domcke, W.; Schlenker, C. W. Control of Excited-State Proton-Coupled Electron Transfer by Ultrafast Pump-Push-Probe Spectroscopy in Heptazine-Phenol Complexes: Implications for Photochemical Water Oxidation. *J. Phys. Chem. C* **2020,** *124*, 9151-9160.
63. Merschjann, C.; Tyborski, T.; Orthmann, S.; Yang, F.; Schwarzburg, K.; Lublow, M.; Lux-Steiner, M.-C.; Schedel-Niedrig, T. Photophysics of Polymeric Carbon Nitride: An Optical Quasimonomer. *Phys. Rev. B: Condens. Matter Mater. Phys.* **2013,** *87*, No. 205204.
64. Zhang, H.; Chen, Y.; Lu, R.; Li, R.; Yu, A. Charge Carrier Kinetics of Carbon Nitride Colloid: A Femtosecond Transient Absorption Spectroscopy Study. *Phys. Chem. Chem. Phys.* **2016,** *18*, 14904-14910.
65. Godin, R.; Wang, Y.; Zwijnenburg, M. A.; Tang, J.; Durrant, J. R. Time-Resolved Spectroscopic Investigation of Charge Trapping in Carbon Nitrides Photocatalysts for Hydrogen Generation. *J. Am. Chem. Soc.* **2017,** *139*, 5216-5224.
66. Corp, K. L.; Schlenker, C. W. Ultrafast Spectroscopy Reveals Electron-Transfer Cascade That Improves Hydrogen Evolution with Carbon Nitride Photocatalysts. *J. Am. Chem. Soc.* **2017,** *139*, 7904-7912.
67. Yang, W.; Godin, R.; Kasap, H.; Moss, B.; Dong, Y.; Hillman, S. A. J.; Steier, L.; Reisner, E.; Durrant, J. R. Electron Accumulation Induces Efficiency Bottleneck for Hydrogen Production in Carbon Nitride Photocatalysts. *J. Am. Chem. Soc.* **2019,** *141*, 11219-11229.
68. Turner, D. W.; Baker, C.; Baker, A. D.; Brundle, C. R. *Molecular Photoelectron Spectroscopy*. Wiley, New York, 1970.
69. Winter, B.; Faubel, M. Photoemission from Liquid Aqueous Solutions. *Chem. Rev.* **2006,** *106*, 1176-1211.
70. Deifallah, M.; McMillan, P. F.; Cora, F. Electronic and Structural Properties of Two-Dimensional Carbon Nitride Graphenes. *J. Phys. Chem. C* **2008,** *112*, 5447-5453.
71. Wei, W.; Jacob, J. Strong Excitonic Effects in the Optical Properties of Graphitic Carbon Nitride g-$C_3N_4$ From First Principles. *Phys. Rev. B: Condens. Matter Mater. Phys.* **2013,** *87*, No. 085202.
72. Srinivasu, K.; Modak, B.; Ghosh, S. K. Porous Graphitic Carbon Nitride: A Possible Metal-Free Photocatalyst for Water Splitting. *J. Phys. Chem. C* **2014,** *118*, 26479-26484.
73. Ma, H.; Feng, J.; Jin, F.; Wei, M.; Liu, C.; Ma, Y. Where Do Photogenerated Holes at the g-$C_3N_4$/Water Interface Go for Water Splitting: $H_2O$ or $OH^-$? *Nanoscale* **2018,** *10*, 16624-15631.
74. Ma, H.; Zhang, X.; Jin, F.; Zhou, H.; Zhang, J.; Ma, Y. Crucial Roles of Triazinic-N=O and C=O Groups in Photocatalytic Water Splitting on Graphitic Carbon Nitride. *J. Mater. Chem. A* **2021,** *9*, 5522-5532.
75. You, P.; Lian, C.; Chen, D.; Xu, J.; Zhang, C.; Meng, S.; Wang, E. Nonadiabatic Dynamics of Photocatalytic Water Splitting on a Polymeric Semiconductor. *Nano Lett.* **2021,** *21*, 6449-6455.
76. Muuronen, M.; Parker, S. M.; Berardo, E.; Le, A.; Zwijnenburg, M. A.; Furche, F. Mechanism of Photocatalytic Water Oxidation on Small $TiO_2$ Nanoparticles. *Chem. Sci.* **2017,** *8*, 2179-2183.
77. Shibata, T.; Kabumoto, A.; Shiragami, T.; Ishitani, O.; Pac, C.; Yanagida, S. Novel Visible-Light-Driven Photocatalyst. Poly(p-phenylene)-Catalyzed Photoreductions of Water, Carbonyl Compounds, and Olefins. *J. Phys. Chem.* **1990,** *94*, 2068-2076.

78. Matsuoka, S.; Kohzuki, T.; Kuwana, Y.; Nakamura, A.; Yanagida, S. Visible-Light-Induced Photocatalysis of Poly(Pyridine-2,5-diyl). Photoreduction of Water, Carbonyl Compounds and Alkenes with Triethylamine. *J. Chem. Soc. Perkin Trans. 2* **1992**, 679-685.
79. Sprick, R. S.; Jiang, J.-X.; Bonillo, B.; Ren, S.; Ratvijtvech, T.; Guiglion, P.; Zwijnenburg, M. A.; Adams, D. J.; Cooper, A. J. Tunable Organic Photocatalysts for Visible-Light-Driven Hydrogen Evolution. *J. Am. Chem. Soc.* **2015,** *137*, 3265-3270.
80. Bai, J.; Wilbraham, L.; Slater, B. J.; Zwijnenburg, M. A.; Sprick, R. S.; Cooper, A. J. Accelerated Discovery of Organic Polymer Photocatalysts for Hydrogen Evolution from Water Through the Integration of Experiment and Theory. *J. Am. Chem. Soc.* **2019,** *141*, 9063-9071.
81. Prentice, A. W.; Zwijnenburg, M. A. Hydrogen Evolution by Polymer Photocatalysts; A Possible Photocatalytic Cycle. *Sust. Energy Fuels* **2021,** *5*, 2622-2632.
82. Prentice, A. W.; Zwijnenburg, M. A. The Role of Computational Chemistry in Discovering and Understanding Organic Photocatalysts for Renewable Fuel Synthesis. *Adv. Energy Mater.* **2021**, No. 2100709.
83. Vyas, V. S.; Haase, F.; Stegbauer, L.; Savasci, G.; Podjaski, F.; Ochsenfeld, C.; Lotsch, B. V. A Tunable Azine Covalent Organic Framework Platform for Visible Light-Induced Hydrogen Generation. *Nature Commun.* **2015,** *6*, No. 8508.
84. Vyas, V. S.; Lau, V. W.; Lotsch, B. V. Soft Photocatalysis: Organic Polymers for Solar Fuel Production. *Chem. Mater.* **2016,** *28*, 5191-5204.
85. Pati, P. B.; Damas, G.; Tian, L.; Fernandes, D. L. A.; Zhang, L.; Pehlivan, I. B.; Edvinsson, T.; Araujo, C. M.; Tian, H. An Experimental and Theoretical Study of an Efficient Polymer Nano-Photocatalyst for Hydrogen Evolution. *Energy Env. Sci.* **2017,** *10*, 1372-1376.
86. Aitchison, C. M.; Sachs, M.; Little, M. A.; Wilbraham, L.; Brownbill, N. J.; Kane, C. M.; Blanc, F.; Zwijnenburg, M. A.; Durrant, J. R.; Sprick, R. S., et al. Structure-Activity Relationships in Well-Defined Conjugated Oligomer Photocatalysts for Hydrogen Production from Water. *Chem. Sci.* **2020,** *11*, 8744-8756.
87. Ehrmaier, J.; Picconi, D.; Karsili, T. N. V.; Domcke, W. Photodissociation Dynamics of the Pyridinyl Radical: Time-Dependent Quantum Wave-Packet Calculations. *J. Chem. Phys.* **2017,** *146*, No. 124304.
88. Xie, J.; Jin, R.; Li, A.; Bi, Y.; Ruan, Q.; Deng, Y.; Zhang, Y.; Yao, S.; Sankar, G.; Ma, D., et al. Highly Selective Oxidation of Methane to Methanol at Ambient Conditions by Titanium Dioxide-Supported Iron Species. *Nature Catal.* **2018,** *1*, 889-896.

**Acknowledgments**

We would like to express our appreciation to Cody W. Schlenker and his students Emily J. Rabe and Kathryn L. Corp at the Department of Chemistry of the University of Washington for extensive stimulating discussions. The close interfacing of experiment and theory described in the present article would not have been possible without their pioneering synthetic and spectroscopic work. We also thank our students Johannes Ehrmaier and Xiang Huang for their essential contributions to the computational work.

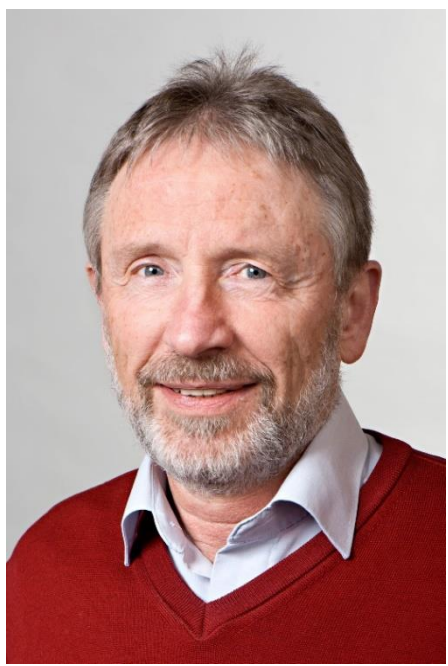

**Wolfgang Domcke** received the PhD degree in Physics from the Technical University of Munich (TUM), Germany, in 1975 and the Habilitation degree from the University of Freiburg, Germany, in 1979. He has been Associate Professor of Theoretical Chemistry at the University of Heidelberg, Germany, and at TUM. In 1996, he became Full Professor of Theoretical Chemistry at the Heinrich-Heine University in Düsseldorf, Germany. He returned to TUM in 1999. Since 2013, he is professor emeritus and member of the TUM Senior Excellence Faculty. In his research he has explored various aspects of nonadiabatic dynamics in molecular spectroscopy and photochemistry.

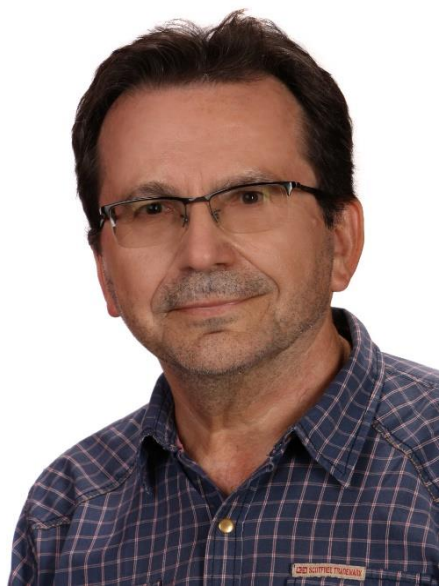

**Andrzej L. Sobolewski** received his MSc degree in Physics from Warsaw University, Poland, in 1977 and was employed as research assistant at the Institute of Physics of the Polish Academy of Sciences (IPPAS) where he received the PhD degree in 1981. He spent two years as Alexander von Humboldt postdoctoral fellow at the Technical University of Munich. Having received the Habilitation degree in 1989, he was promoted to professor at IPPAS. His research interests include, among others, computational investigations of elementary photochemical processes, e.g. excited-state proton and electron transfer reactions, and pathways of radiationless energy relaxation in molecular systems.